# Effect of irradiation by He$^+$ and Ga$^+$ ions on 2D-exciton susceptibility of the InGaAs/GaAs quantum-well structures

Yu.V. Kapitonov[*,1], P.Yu. Shapochkin[1], Yu.V. Petrov[1], Yu.P. Efimov[1], S.A. Eliseev[1], Yu.K. Dolgikh[1], V.V. Petrov[1], V.V. Ovsyankin[1]

[1] Saint Petersburg State University, Ulyanovskaya 1, 198504 St. Petersburg, Russia

* Corresponding author: e-mail kapiton22@gmail.com, Phone: +7 812 428 45 66, Fax: +7 812 428 72 40

The effect of irradiation by 30-keV Ga$^+$ and 35-keV He$^+$ ions (in relatively small doses) on the excitonic reflectivity spectra of single InGaAs/GaAs quantum-well structures is studied. It is found that the irradiation results in decreasing intensity and broadening of the excitonic resonances in the reflectivity spectra for all the doses. It is shown that these changes are not related to a decrease of the exciton transition oscillator strength and, therefore, to the irradiation-induced destruction of the excitonic states, but can be rather ascribed to a common cause, namely, to inhomogeneous broadening of the excitonic resonances proportional to the exposure dose. A tentative model of the irradiation-induced broadening is considered, with the mechanism of the process being a consequence of scattering of the 2D excitons by structural defects associated with Ga(In) and As vacancies arising upon collisions of the high-energy ions with regular atoms of the crystal structure. The model is used to compare experimental dependence of efficiency of the Ga$^+$-ion-induced broadening on distance of the quantum well from the irradiated surface with a similar dependence calculated using the Monte-Carlo technique. A discrepancy between the results of simulation and experimental data is discussed.

**1 Introduction** In [1], we have proposed and realized a method of producing resonant diffraction optical elements on the basis of 2D-exciton mirrors by addressable spatial modulation of the exciton resonance width. The modulation was achieved by growing the quantum-well (QW) structures on substrates with a preliminary modified surface structure. The modification implements addressable irradiation of the substrate by Ga$^+$ Focused Ion Beam (FIB). The irradiation caused formation of structural defects in the substrate and deposition of carbon nano-layers over the irradiated regions. Spatial resolution of this method meets the requirements of the diffraction optics and is limited by the fact that the growth rate of the epitaxial layers is ten times lower than the lateral spreading rate of structural defects initiated by modification of the substrate surface. A technological drawback of this method is related to uncontrollable contamination of the epiready substrate at the stage of surface modification, which hampers growth of the high-quality epitaxial structures.

This paper aims at searching a method of spatial modulation of optical susceptibility of the 2D-exciton mirrors free from the above physical and technological limitations. As a promising alternative, we consider possibility to create additional channels of scattering for the 2D-excitons by introducing structural defects into the ready QW structures by irradiating it with high-energy ions. This possibility is based on the well-known fact that structural defects (vacancies, interstitial atoms, and their associates) may be produced due to inelastic scattering of high-energy ions propagating in crystal structures.

Up-to-date equipment allows one to obtain 30 keV Ga$^+$ ion beams focused to a less than 10 nm spot [2] and 35 keV He$^+$ ions focused to a 1 nm spot [3]. Still, in spite of remarkable abilities of the present-day technology, regularities of the effect of ion bombardment on optical susceptibility of the QW 2D-excitons and their correlation with known regularities of the irradiation-induced defect formation, as far as we know, has not been studied. One of the goals of this paper is to study these issues.

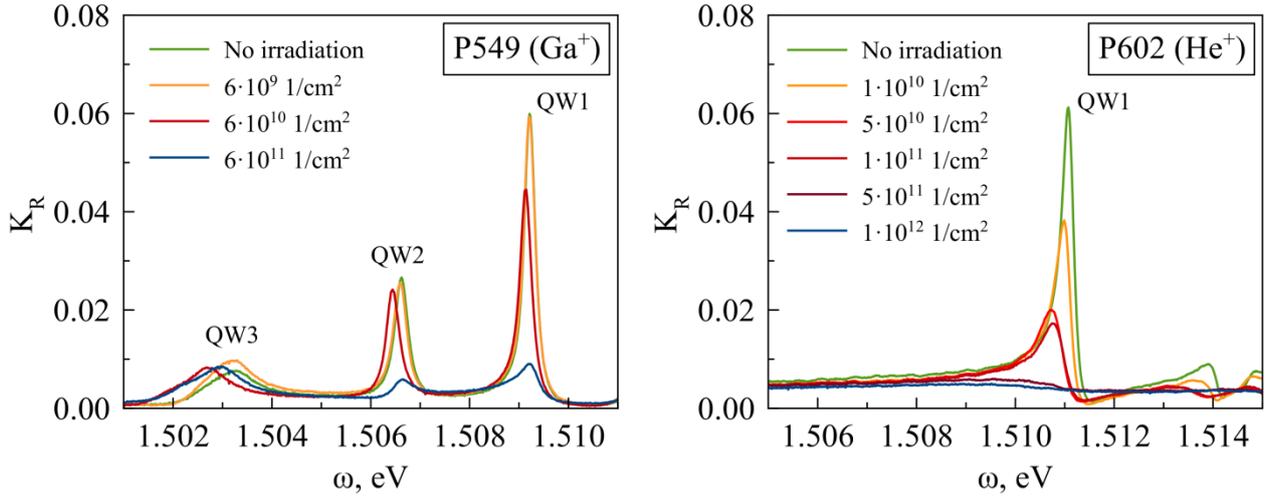

**Figure 1** Reflectivity spectra of the samples P549 and P602 in the areas with different ion-irradiation doses and with no irradiation.

## 2 Experimental

**2.1 Sample preparation** We studied samples with InGaAs/GaAs single quantum wells grown in an EP-1302 Molecular Beam Epitaxy (MBE) machine. To investigate the effects of Ga$^+$ ion irradiation, we used the sample P549 with 3 narrow quantum wells QW1, QW2, and QW3, separated by 50 nm GaAs barriers. The effects of the He$^+$ ion irradiation was studied using the sample P602 with a narrow QW and a cap layer 60 nm thick. The thicknesses of the layers and their distances from the surface $h$ found from the RHEED data are given in Table 1.

**Table 1** Parameters of quantum wells.

|  | P549 | | | P602 |
|---|---|---|---|---|
|  | QW1 | QW2 | QW3 | QW1 |
| Medium | In(3%)GaAs | | InAs | In(1.5%)GaAs |
| Width, nm | 3.2 | 4.2 | 0.16 | 4.5 |
| $h$, nm | 176 | 230 | 280 | 60 |
| Ions | 30 keV Ga$^+$ | | | 35 keV He$^+$ |
| $\alpha$, µeV·cm$^2$ | 3.4×10$^{-10}$ | 2.7×10$^{-10}$ | - | 2.2×10$^{-9}$ |

Irradiation of the sample P549 by the 30 keV Ga$^+$ ions was performed using the FIB-system Zeiss Crossbeam 1540XB supplied with the external scan generator Raith Elphy PLUS. Areas of 600×600 µm were uniformly irradiated on the sample at a current of 1 pA. The dose needed for etching the GaAs by the 30 keV Ga$^+$ beam [4] to a depth of 1 nm lies in the range of 10$^{14}$ 1/cm$^2$. The doses used in our experiments were much lower. We exposed three areas with the doses 6×10$^9$, 6×10$^{10}$, and 6×10$^{11}$ 1/cm$^2$.

Irradiation of the sample P602 by the 35 keV He$^+$ ions was performed using the Helium Ion Beam Microscope ORION supplied with the external scan generator Nanomaker. Areas of 300×150 µm were uniformly irradiated on the sample at a current of 0.15 pA with a step of the filling raster being 50 nm. Irradiation of GaAs by He$^+$ ions with a dose exceeding 10$^{16}$ 1/cm$^2$ was accompanied by appearance of bubbles and, with further increase of the dose, by blistering [5]. The doses used in our experiments were much smaller (1×10$^{10}$, 5×10$^{10}$, 1×10$^{11}$, 5×10$^{11}$, and 1×10$^{12}$ 1/cm$^2$).

**2.2 Optical measurements** As the main method of studying excitonic properties of the irradiated and non-irradiated epitaxial structures, we used reflection spectroscopy in the Brewster configuration [1,6] based on the theory developed in [7]. Theoretical consideration of the reflection signal leads to the following expression for spectral behavior of the exciton mirror reflectivity $K_R(\omega,D)$:

$$K_R(\omega,D) = \frac{\Gamma_R^2}{(\omega-\omega_0)^2 + (\Gamma_R + \Gamma_2 + \Gamma_2^* + \Delta\Gamma_2^*(D))^2},$$

where $2\hbar\Gamma_R$ is the radiative width of the exciton resonance, $2\hbar\Gamma_2$ and $2\hbar\Gamma_2^*$ are its homogeneous and inhomogeneous width, and $2\hbar\Delta\Gamma_2^*(D)$ is the additional inhomogeneous broadening arising due to irradiation of the sample by ions with the dose $D$, and $\omega_0$ is the exciton resonance frequency.

Figure 1 shows the reflectivity spectra at $T = 9.5$ K for the areas of the samples with different exposure doses and of the reference non-irradiated area. Resonances indicated in the figure correspond to HH-excitons of the appropriate QWs. To analyze the effect of ion irradiation, the reflectivity spectra of the irradiated spots were compared with those of non-irradiated spots lying nearby. The experimental error was estimated from the measurements made at several points of the irradiated region.

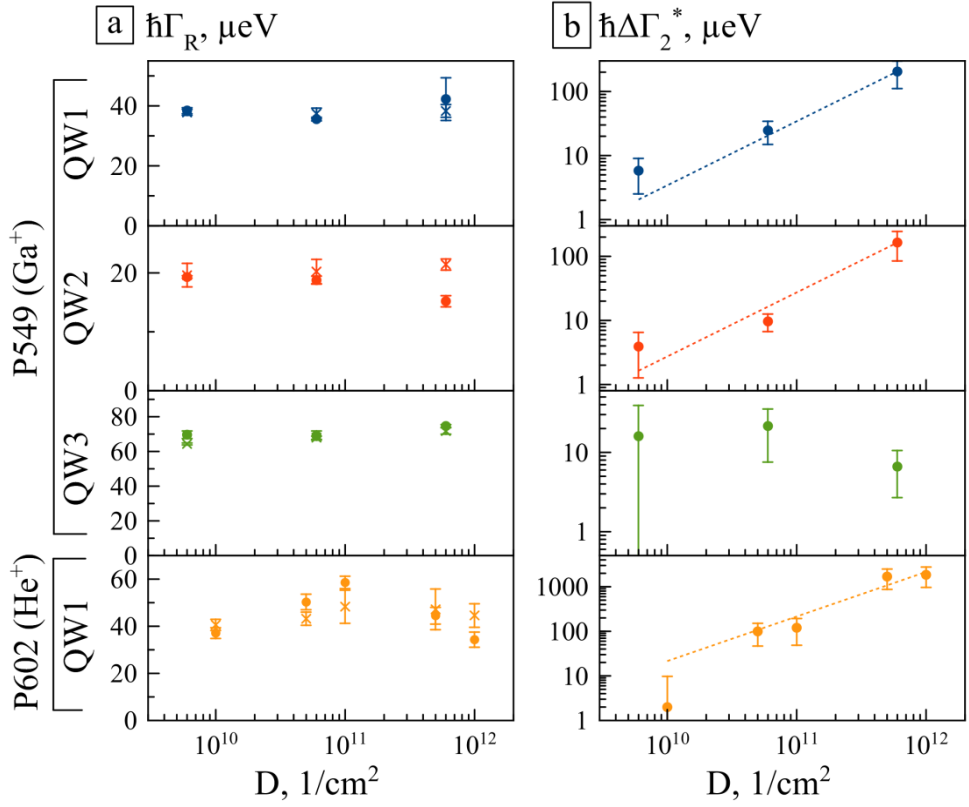

**Figure 2** The radiative $\hbar\Gamma_R$ width (a) prior to (x) and after (o) the ion irradiation, the additional inhomogeneous broadening $\hbar\Delta\Gamma_2^*$ (b), and the linear approximation $\alpha D$ (dashed line) as a function of the irradiation dose $D$.

Fitting of the experimental data using the transfer matrix method allowed us to extract values of $\hbar\Gamma_R$ and $\hbar\Delta\Gamma_2^*(D)$ (Fig.2). The $\hbar\Gamma_R(D)$ dependences presented in Fig. 2 (a) show that the ion irradiation does not noticeably affect the oscillator strengths of the exciton resonances. Fig. 2 (b) also shows linear approximations of the dose dependence $\hbar\Delta\Gamma_2^*(D) = \alpha D$. The values of the irradiation-induced broadening factors $\alpha$ thus obtained are given in Table 1. Analysis of the P549 QW3 resonance broadening was hampered because of a large initial width of the resonance.

Practically total absence of any non-resonant background Rayleigh scattering in the reflectivity spectra shows that the ion irradiation, in this case, changes only resonant properties of the sample, leaving optical quality of its surface intact.

**3 Discussion** Analysis of the quantities $\alpha(h)$ was based on the Monte-Carlo simulation of ion scattering in the sample using the program TRIM [8]. The ion beam was assumed incident at normal direction, the GaAs crystal density was taken to be 5.32 g/cm$^3$, the displacement energy was taken the same for Ga and As $E_d$ = 9.5 eV as in [9]. The simulation was performed with allowance for secondary ion cascades. We calculated the total Ga and As vacancy generation yield $V(h)$ as a function of the distance of the layer from the surface $h$. The product of $V(h)$ and the exposure dose $D$ allows one to calculate the density of vacancies at a given depth $h$.

Figure 3 shows the $V(h)$ profile for irradiation by the 30 keV Ga$^+$ and 35 keV He$^+$ ions. One can notice a larger penetration depth and smaller damage capacity of the He$^+$ ions, which can be ascribed to their smaller mass compared with atoms of the GaAs lattice. The structural defects arising in the QW layer lead to exciton scattering and additional inhomogeneous broadening of the exciton resonances. At low defect densities, the additional inhomogeneous broadening is proportional to this density and, therefore, to the exposure dose, which was confirmed experimentally (Fig.2 (b)).

Figure 3 (b) shows dependence of the vacancy generation yield $V(r)$ on the distance from the axis of the focused ion beam $r$ for the layer lying at the depth $h$ = 20 nm. It has been shown experimentally that the InGaAs/GaAs QWs with such a thin cap layer retain their high quality [10]. The product of $V(h)$ and the exposure dose $D$ allows one to calculate the density of vacancies at a given depth $h$ and distance from the beam axis $r$. It is seen from Fig. 3 (b) that, for sufficiently tight beam focusing, one can achieve a sub-100-nm resolution for spatial modulation of the QW defect density. It is noteworthy that the choice in favor of the $Ga^+$ or $He^+$ ions can be made based or requirements to the ratio between resolution and contrast of the modulation and with allowance for the distance of the QW from the surface.

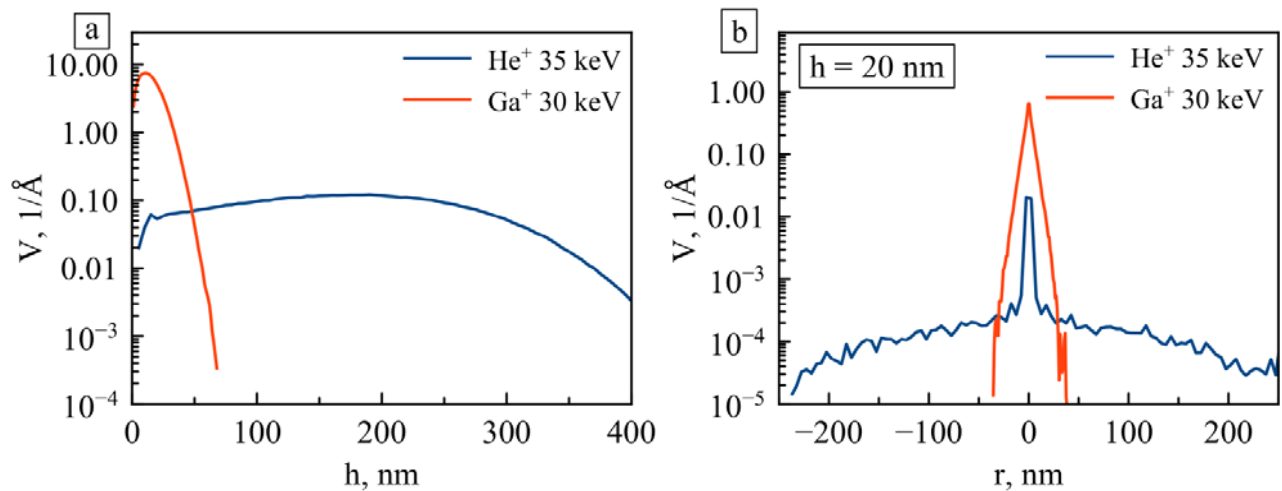

**Figure 3** The vacancy generation yield $V$ as a function of the depth $h$ for the case of uniform irradiation (a) and as a function of the distance from the beam axis $r$ for the layer at the depth $h$=20 nm for the case of focused beam (b).

The Monte-Carlo simulation predicts a negligibly small vacancy density at the depth of the P549 QW2 and QW3 layers for the 30 keV $Ga^+$ ion irradiation. However, we have found experimentally the possibility of irradiation-induced modulation of their optical characteristics. In addition, the values of irradiation-induced broadening factors $\alpha$ obtained for P549 QW1 and QW2 are close to each other in spite of the fact that the wells are separated by ~50 nm of GaAs.

There are possible reasons for the discrepance between the results of simulation and our experimental data: the penetration depth for these ions may appear to be substantially larger because of the channeling effects in the crystal ignored in Monte-Carlo calculations [9,11], and the broadening may be caused by scattering on free carriers produced by some donor or acceptor states related to irradiation-induced defects of the structures [12,13]. For refinement of reasons of the discrepancy additional experiments are needed, e.g., with the quantum wells at larger distances from the surface. As for analysis of resolution of the spatial modulation and its comparison with the results of simulation, it can be made by creating periodic diffraction structures as in [1].

**3 Conclusions** In this paper, we demonstrate the effect of irradiation by 30 keV $Ga^+$ and 35 keV $He^+$ ions on the exciton reflectivity spectra of single InGaAs/GaAs quantum wells. The ion irradiation leads to appearance of additional structural defects in the QW layer, which, in turn, additionally broadens the exciton resonances.

We have found experimentally that the exposure dose for the 30 keV $Ga^+$ and 35 keV $He^+$ ion irradiation needed for substantial broadening of resonances of the QWs located at the depth smaller than 250 nm is smaller than the exposure doses capable of sputtering the sample by several orders of magnitude. Irradiation of the sample with these doses does not change the resonance oscillator strength, with the inhomogeneous broadening of the resonance increasing in proportion with the irradiation dose.

For the $Ga^+$ ion irradiation, the efficiency of broadening of the exciton resonance was measured for quantum wells lying at different depth. The $Ga^+$ ion-irradiation-induced generation of vacancies in GaAs was simulated using the Monte-Carlo technique. The depth profiles calculated in this way essentially differ from the experimental data. Several possible reasons for this discrepancy are considered.

**Acknowledgements** The reported study was partially supported by RFBR, research project No. 14-02-31617 mol_a. The authors also acknowledge Saint-Petersburg State University for research Grants No. 11.38.67.2012 and 11.0.66.2010. This work has been partially funded by OPTEC Zeiss grant and carried out on the equipment of the SPbU Resource Centers "Nanophotonics" (photon.spbu.ru) and "Nanotechnology" (nano.spbu.ru).

**References**

[1] Yu. V. Kapitonov, M. A. Kozhaev, Yu. K. Dolgikh, S. A. Eliseev, Yu. P. Efimov, P. G. Ulyanov, V. V. Petrov, and V. V. Ovsyankin, Phys. Status Solidi B **250**, No.10, 2180 (2013).

[2] C. A. Volkert and A. M. Minor, MRS Bulletin **32**, 389 (2007).

[3] R. Hill, J.A. Notte, L. Scipioni, Advances in Imaging and Electron Physics **170**, 65 (2012).

[4] R. Menzel, T. Bachmann, and W. Wesch, Nuclear Instruments and Methods in Physics Research B **148** 450 (1999).

[5] D. M. Follstaedt, S. M. Myers, J. C. Barbour, G. A. Petersen, J. L. Reno, L. R. Dawson, and S. R. Lee, Nuclear Instruments and Methods in Physics Research B **160** 476 (2000).

[6] S. V. Poltavtsev and B. V. Stroganov, Phys. Solid State **52**, 1899 (2010).

[7] M. G. Benedict and E. D. Trifonov, Phys. Rev. A **38**, 2854 (1988).

[8] J. F. Ziegler, M. D. Ziegler, J. P. Biersack, Nuclear Instruments and Methods in Physics Research B **268** 1818 (2010).

[9] R. Menzel, K. Gärtner, W. Wesch, and H. Hobert, Journal of Applied Physics **88**, 5658 (2000).

[10] S. M. Wang, J. V. Thordson, T. G. Andersson, S. Jiang, L. X. Yang, and S. C. Shen, Applied Physics Letters **65**, 336 (1994).

[11] M. Gal, M. C. Wengler, S. Ilyas, I. Rofii, H. H. Tan, and C. Jagadish, Nuclear Instruments and Methods in Physics Research B **173**, 528 (2001).

[12] E. H. Linfield, D. A. Ritchie, G. A. C. Jones, J. E. F. Frost, and D. C. Peacock, , Semicond. Sci. Technol. **5**, 385 (1990).

[13] E. H. Linfield, G. A. C. Jones, D. A. Ritchie, and J. H. Thompson, , Semicond. Sci. Technoi. **8**, 415 (1993).